\documentclass{article}

\PassOptionsToPackage{numbers, compress}{natbib}

\usepackage[final]{ewrl_2024}

\usepackage[utf8]{inputenc} 
\usepackage[T1]{fontenc}    
\usepackage{hyperref}       
\usepackage{url}            
\usepackage{booktabs}       
\usepackage{amsfonts}       
\usepackage{nicefrac}       
\usepackage{microtype}      
\usepackage{xcolor}         
\usepackage{graphicx}
\usepackage{array}
\usepackage{amsmath}
\usepackage{multirow}
\usepackage{geometry}
\usepackage{booktabs}

\usepackage[utf8]{inputenc}
\usepackage{pgfplots}
\pgfplotsset{compat=1.9}
\usepgfplotslibrary{groupplots}
\usepgfplotslibrary{fillbetween}
\usepgfplotslibrary{colorbrewer}

\pgfplotsset{
    myplotstyle/.style={
    legend style={draw=none, font=\small},
    legend cell align=left,
    legend pos=north east,
    ylabel style={align=center, font=\bfseries\boldmath},
    xlabel style={align=center, font=\bfseries\boldmath},
    x tick label style={font=\bfseries\boldmath},
    y tick label style={font=\bfseries\boldmath},
    scaled ticks=false,
    every axis plot/.append style={thick},
    },
}

\title{Environment Complexity and Nash Equilibria in a Sequential Social Dilemma}
\author{%
  Mustafa Yasir \\
  The Alan Turing Institute\\
  \texttt{myasir@turing.ac.uk} \\
  \And
  Andrew Howes \\
  University of Exeter\\
  \texttt{andrew.howes@exeter.ac.uk} \\
  \And
  Vasilios Mavroudis \\
  The Alan Turing Institute\\
  \texttt{vmavroudis@turing.ac.uk} \\
  \And
  Chris Hicks \\
  The Alan Turing Institute\\
  \texttt{c.hicks@turing.ac.uk} \\
}

\begin{document}

\maketitle

\begin{abstract}
  Multi-agent reinforcement learning (MARL) methods, while effective in zero-sum or positive-sum games, often yield suboptimal outcomes in general-sum games where cooperation is essential for achieving globally optimal outcomes. Matrix game social dilemmas, which abstract key aspects of general-sum interactions, such as cooperation, risk, and trust, fail to model the temporal and spatial dynamics characteristic of real-world scenarios. In response, our study extends matrix game social dilemmas into more complex, higher-dimensional MARL environments. We adapt a gridworld implementation of the Stag Hunt dilemma to more closely match the decision-space of a one-shot matrix game while also introducing variable environment complexity. Our findings indicate that as complexity increases, MARL agents trained in these environments converge to suboptimal strategies, consistent with the risk-dominant Nash equilibria strategies found in matrix games. Our work highlights the impact of environment complexity on achieving optimal outcomes in higher-dimensional game-theoretic MARL environments. 
\end{abstract}

\section{Introduction}
Multi-agent reinforcement learning (MARL) is concerned with training agents to maximise individual or shared rewards in environments with multiple concurrent learners. A critical aspect of MARL is the agent motive, which can be categorised as competitive (zero-sum), purely cooperative (positive-sum), or mixed (general-sum). MARL methods have shown considerable success both in competitive environments \citep{Jaderberg_2019, Vinyals2019GrandmasterLI} and in purely cooperative environments \citep{10.5555/3237383.3238080, 10.5555/3454287.3454752}. In these settings, agent goals are either directly opposed or aligned, and learning paradigms like self-play, which train agents to compute best responses to rational opponents, empirically converge to stable strategies, i.e., Nash equilibria, that correspond to optimal outcomes. However, Nash equilibria in general-sum games often coincide with suboptimal outcomes, rendering the direct application of such methods ineffective for finding desirable solutions \citep{10.5555/3237383.3237408, dafoe2020open, lerer2018maintaining}. Instead, in many general-sum games, optimal outcomes are often only achievable through cooperation—by explicitly favouring collective gains over stable individual rewards. 

Matrix game social dilemmas (MGSDs), a class of general-sum games, represent situations where individual rationality may lead to collective irrationality. These dilemmas are widely used to explore cooperation in various disciplines \citep{Schelling1960,Smith1976EvolutionAT}, including MARL, where numerous methods have been proposed to induce cooperation among learning agents \cite{Du_Leibo_Islam_Willis_Sunehag_2023}. While MGSDs are valuable for abstracting key aspects of real-world multi-agent interactions—such as cooperation, risk, and trust—they lack the complexity of real-world temporal and spatial dynamics. Unlike real-world actions, matrix game actions are binary, i.e., limited to cooperate or defect choices, and do not model extended timeframes \citep{Leibo2017MultiagentRL}. Consequently, recent works have looked at extending MGSDs to higher-dimensional MARL environments in order to more accurately model real-world scenarios (see Section \ref{sec:rel work}).

In this study, we focus on an extension of the Stag Hunt MGSD to a gridworld MARL environment, and explore how unique environment dynamics—absent in the matrix game—affect the learnt strategies of independent MARL agents. Specifically, we introduce eight novel variants of the gridworld Stag Hunt, each incoporating a different degree of randomisation across various dynamics, which we model as environment complexity. This research direction is motivated by the inherent limitations of MGSDs, where modifications to a game are typically confined to changes in payoff structure. Extending these games into higher-dimensional environments offers a novel opportunity to examine how additional factors, such as environment dynamics, influence cooperative behaviours. 

Furthermore, our inquiry is supported by studies on randomness in evolutionary social dilemmas \cite{Szolnoki_2009}, along with empirical research highlighting its impact on cooperation \citep{Baker2002Teaching, Yang2007The}. A key finding is that multiple variants of the gridworld Stag Hunt can map to the same Stag Hunt MGSD; however, even minor variations in their complexity can promote convergence to vastly different strategies. Moreover, our work advances the development of game-theoretic MARL environments, more accurately reflecting real-world conditions, where environment dynamics may play a significant role in shaping general-sum multi-agent interaction. In summary, we present the following key contributions:
\begin{itemize}
    \item We adapt a multi-agent gridworld implementation of the Stag Hunt MGSD to match the decision space of a one-shot matrix game and create eight unique environment variants, each with a specified level of complexity.
    \item We train independent MARL agents in each of these environment variants and observe distinct empirical convergence patterns to either the suboptimal Nash equilibrium strategy or a mixed strategy. 
    \item We demonstrate that environment complexity promotes systematic convergence to the suboptimal Nash equilibrium strategy, despite the established viability of more optimal strategies.
    \item  We conduct an empirical game-theoretic analysis on our trained policies, showing that certain gridworld Stag Hunt variants can be mapped to MGSDs with Stag Hunt payoffs, and can be modelled as Sequential Social Dilemmas. 
\end{itemize}

\section{Related Work}
\label{sec:rel work}
Addressing the viability of MGSDs in modelling key aspects of multi-agent interaction, numerous studies have proposed higher-dimensional game-theoretic MARL environments to explore cooperation under more realistic settings. The Coin Game, a widely used gridworld environment based on repeated MGSDs, has been used to examine cooperative MARL methods by \citet{10.5555/3237383.3237408}, and the Melting Pot suite combines various MGSDs in gridworld formats to benchmark MARL algorithms \citep{leibo2021scalable}. Recently, \citet{10.5555/3635637.3662955} highlight the challenges in adapting matrix games to gridworld environments by demonstrating the shortcomings of the Coin Game. Relevant to our focus on the Stag Hunt, is the Level-Based Foraging environment \cite{christianos2020shared}, a gridworld foraging game which targets multi-agent cooperation and coordination, and a study by \citet{8531285}, which successfully demonstrates enhanced learning in a Minecraft-based stag hunt through gridworld abstraction. Although existing works provide useful game-theoretic environments, primarily for proposing new MARL methods for repeated and partially observable settings, we simplify the learning problem to one-shot and fully observable games; to focus on modifying environment dynamics, whilst maintaining the relation between our environment and its matrix game formulation. To this end, we leverage an existing game-theoretic MARL environment \cite{nesterov-rappoport_2022, Peysakhovich2017ProsocialLA} , based on the Stag Hunt MGSD, chosen for its overlap with existing environments and accessibility of manipulating game dynamics, which was briefly explored by \citet{Peysakhovich2017ProsocialLA}.

\section{Background}

\subsection{Matrix Game Social Dilemmas}
\label{sec:MGSD}
A matrix game is a formal representation of strategic interactions between two players, where each player’s decision affects the other’s outcome. The game is represented by a matrix in which each cell details the outcomes (or payoffs) for the players based on their combined choices. Players choose a strategy without knowledge of the other’s simultaneous decision, and the combination of their choices leads to a specific payoff for each. In matrix game social dilemmas (MGSDs), the two actions, or strategies, available to each player are \textit{`cooperate`} and \textit{`defect`}, and the exact payoffs for each player are given by the values $\{R, P, S, T \} \in \mathbb{R}$, detailed in Table \ref{tab:MGSDmatrix}. Formally, a matrix game is a social dilemma when its four payoffs satisfy the following inequalities:

\begin{equation}
\label{eq:MGSDineq}
     R > P, R > S, 2R > T+S, \text{ and either: } T>R, \text{or, } P>S. 
\end{equation}

\begin{table}[h]
\centering
\caption{Payoff table for a matrix game social dilemma, where ($X,Y$) denotes the payoff for the given row and column player, respectively.}
\label{tab:MGSDmatrix}
\begin{tabular}{@{}ccccc@{}}
\toprule
 & & \multicolumn{2}{c}{Player 2} \\ \cmidrule(l){3-4}
 & &  \( Cooperate \) &  \( Defect \) \\ \midrule
\multirow{2}{*}{Player 1} & \( Cooperate \) & \( (R, R) \) & \( (S, T) \) \\
                          & \( Defect \) & \( (T, S) \) & \( (P, P) \) \\
\bottomrule
\end{tabular}
\end{table}

Central to our work, is the MGSD, Stag Hunt \cite{Rooij2007TheSH}; where the \textit{`cooperate`} and \textit{`defect`} actions correspond to \textit{`hunt`} and \textit{`forage`}, respectively. Conceptually, this represents a scenario where players are tasked with either hunting a stag or foraging a plant. A bilateral decision to hunt offers the highest payoff, $h$, to both players, but carries the risk of being mauled by the stag, yielding the lowest possible reward, $m$, to a player hunting unilaterally. In mauling scenarios, the opposing player may receive a unique foraging reward $f^* \geq f$. Formally, the Stag Hunt is characterised by Table \ref{tab:payoffMatrix} and corresponds to an MGSD where $R=h$, $P=f$, $S=m$ and $T=f^*$.

\begin{table}[h]
\centering
\caption{Payoff table for the generalised 2-player stag hunt, where $h > f^* \geq f > m$.}
\label{tab:payoffMatrix}
\begin{tabular}{@{}ccccc@{}}
\toprule
 & & \multicolumn{2}{c}{Player 2} \\ \cmidrule(l){3-4}
 & &  \( Hunt \) &  \( Forage \) \\ \midrule
\multirow{2}{*}{Player 1} & \( Hunt \) & \( (h, h) \) & \( (m, f^*) \) \\
                          & \( Forage \) & \( (f^*, m) \) & \( (f, f) \) \\
\bottomrule
\end{tabular}
\end{table}

\subsection{Nash Equilibria}
\label{sec:NE}
Nash Equilibria, a fundamental concept in game theory introduced by \citet{Nash_1950}, refers to a situation in which each player in a game makes an optimal choice considering the choices of the other players, and no player has anything to gain by changing only their own strategy unilaterally. Formally, a Nash Equilibrium in a game with $n$ players is defined as a strategy profile $(s_1^*, s_2^*, \dots, s_n^*)$ such that for each player $i$, the strategy $s_i^*$ is a best response to the strategies $s_{-i}^*$ chosen by the other players. This can be expressed as:
\begin{equation}
s_i^* \in \arg\max_{s_i} u_i(s_i, s_{-i}^*), \quad \forall i \in \{1, 2, \dots, n\}
\end{equation}
where $s_i$ represents the strategy chosen by player $i$, $s_{-i}^*$ represents the strategies chosen by all other players except $i$, and $u_i(s_i, s_{-i})$ is the payoff function for player $i$. In the Stag Hunt, players are faced with the decision between hunting a stag or foraging for plants, i.e., $s_i \in \{Hunt, Forage\}$, and the game is particularly valuable for containing two Nash Equilibria which reflect different resolutions of its inherent social dilemma:

\paragraph{Payoff-Dominant Equilibrium $(Hunt, Hunt)$:} If both players choose to hunt stag, neither has an incentive to deviate unilaterally since doing so would result in a lower payoff. If one player switches to foraging while the other continues to hunt, the deviating player’s payoff decreases from $h$ (under successful cooperation) to $f$ or $f^*$, both of which are less than $h$ under the game’s assumptions. Thus, $(Hunt, Hunt)$ is a Nash Equilibrium. This outcome is payoff dominant because it results in the highest rewards for both players, i.e., both receive $h$, which is greater than any other payoff in the game. 

\paragraph{Risk-Dominant Equilibrium $(Forage, Forage)$:} Similarly, if both players choose to forage, deviating to hunting is disadvantageous. A unilateral switch would result in a payoff of $m$, as the lone hunter would face the risk of mauling, which is the worst outcome. Hence, $(Forage, Forage)$ also constitutes a Nash Equilibrium. Although this equilibrium yields a lower payoff, $f$, than mutual stag hunting, it is considered risk dominant because it minimises the risk associated with the potential non-cooperation of the other player. 

\subsection{Markov Games}
\label{sec:markovGames}
We model the MARL adaptation of Stag Hunt as a two-player Markov Game, also known as a Stochastic Game \cite{Shapley_1953}; a natural extension of Markov Decision Processes to multi-agent settings \cite{10.5555/3091574.3091594}. A Markov game for two players is a tuple $\langle S, {A}, {R}, P, \gamma \rangle$, where $S$ is the set of states describing the environment, $A$ is the combined action space across all players, given by $A := A_1 \times A_2$, where the $A_i$ is is the set of actions available to player $i$, $R$ is the reward function, given by $R := R_1 \times R_2$, where $R_i : S \times A \times S \rightarrow \mathbb{R}$ is the reward function for player $i$, which maps a state and action tuple (including the resulting state) to a real number and defines the reward that player $i$ receives after all players choose their actions. $P : S \times A \times S \rightarrow \Delta(S)$ is the state transition probability function, and $\gamma$ is the discount factor, which is typically constrained to $0 \leq \gamma \leq 1$. 

In a two-player Markov Game, each player chooses a policy that determines their action based on the current state of the game. The joint policy vector  $\vec{\pi}$  combines the individual policies of each player, where  $\vec{\pi} = (\pi_1, \pi_2)$. For player $i$ , the value function under this joint policy,  $V_i^{\vec{\pi}}$ , is defined as the expected sum of discounted rewards when both players adhere to their respective parts of  $\vec{\pi}$ . The formal definition is:
\begin{equation}
V_i^{\vec{\pi}}(s) = \mathbb{E}_{\vec{a}_t \sim \vec{\pi}, s_{t+1} \sim P(s_t, \vec{a}_t)}\left[\sum_{t=0}^{\infty} \gamma^t R_i(s_t, \vec{a}_{t}, s_{t+1}) \bigg| s_0=s\right]
\end{equation}
where $s_t$ is the state at time  $t$, $\vec{a}_{t}$ is the joint actions taken by Player 1 and Player 2 at time $t$, determined by their policies $\pi_1$ and  $\pi_2$, respectively, $s_{t+1}$ is the state resulting from the actions $\vec{a}_{t}$, $R_i(s_t, \vec{a}_{t}, s_{t+1})$ is the reward received by player $i$ after the actions are taken and the state transitions, and $\gamma$ is the discount factor.

\subsection{Sequential Social Dilemmas}
We also use the Sequential Social Dilemma (SSD) model, introduced by \citet{Leibo2017MultiagentRL}, which characterises when a Markov Game contains embedded MGSDs, thus extending the MGSD model to temporally and spatially extended settings. Formally, an SSD is a tuple $\langle\mathcal{M}, \Pi^C, \Pi^D\rangle$, where $\mathcal{M}$ is a Markov Game with state space $\mathcal{S}$, and, $\Pi^C$ and $\Pi^D$, are disjoint sets of policies that implement cooperative and defective strategies, respectively. These policy sets are defined by selecting a metric, $\alpha: \Pi \rightarrow \mathbb{R}$, and threshold values $\alpha_c$ and $\alpha_d$ such that $\alpha(\pi) < \alpha_c \iff \pi \in \Pi_C$ and $\alpha(\pi) > \alpha_d \iff \pi \in \Pi_D$. For a given Markov Game, $\mathcal{M}$, and state, $s \in \mathcal{M}$, we can construct an empirical payoff matrix, $(R(s), P(s), S(s), T(s))$, induced by the outcomes of policies, $(\pi^C \in \Pi^C, \pi^D \in \Pi^D)$, via the following equations: $R(s) := V_{1}^{\pi^C, \pi^C}(s) = V_{2}^{\pi^C, \pi^C}(s)$, $P(s) := V_{1}^{\pi^D, \pi^D}(s) = V_{2}^{\pi^D, \pi^D}(s)$, $S(s) := V_{1}^{\pi^C, \pi^D}(s) = V_{2}^{\pi^D, \pi^C}(s)$, and $T(s) := V_{1}^{\pi^D, \pi^C}(s) = V_{2}^{\pi^C, \pi^D}(s)$. Then, $\mathcal{M}$ is an SSD when there exist states $s \in S$ and policies $(\pi^C \in \Pi^C, \pi^D \in \Pi^D)$ such that the induced empirical payoff matrix satisfies the five MGSD inequalities in Equation \ref{eq:MGSDineq}. Hence, a Markov Game is an SSD when it can be mapped by empirical game-theoretic analysis \cite{Wellman2006MethodsFE} to an MGSD. 

\subsection{Proximal Policy Optimisation}
\label{sec:PPO}
Proximal Policy Optimisation (PPO) is a widely used policy gradient method for single-agent reinforcement learning, introduced by \citet{schulman2017proximal}. The core idea behind PPO is the use of a clipped surrogate objective function to prevent the policy from deviating too far from the current policy during updates. Through its extension to the multi-agent paradigm, PPO has proven effective in numerous MARL environments, including those that target cooperation \cite{Yu2021TheSE}. 

\section{Methods}

\subsection{Gridworld Stag Hunt}
We adapt a gridworld implementation of the Stag Hunt matrix game, initially developed by \citet{nesterov-rappoport_2022} and \citet{Peysakhovich2017ProsocialLA}. The selected environment, built on the PettingZoo \citep{terry2021pettingzoo} and Gym \citep{1606.01540} frameworks, simulates a 5x5 grid game featuring two agents—red and blue, visualised in Appendix \ref{app:ENV}. Each episode begins with placing both agents, along with non-agent entities: one stag and two plants, on distinct cells of the grid. The agents and the stag can move in any of the four cardinal directions or remain stationary at every timestep. The state of the environment is fully observable for each agent, with integer arrays representing the coordinates of all agents and entities on the grid. Similarly, the action space is represented by five unique integers, corresponding to the four cardinal directions and the option to remain stationary. Plants remain stationary within an episode and are randomly re-spawned at every episode start. Individual agent rewards are dependent on the actions of both agents and are determined at the end of every timestep as follows: 
\begin{itemize}
    \item Agents receive an individual hunting reward, $h$, by jointly occupying the stag's grid cell.
    \item An agent receives a foraging reward, $f$, by independently, or jointly, occupying a plant's grid cell.
    \item An agent receives a mauling reward, $m$, by independently occupying the stag's grid cell.
\end{itemize}

The rewards values assigned satisfy the generalised two-player stag hunt inequality in Table \ref{tab:payoffMatrix}, with values set at $h=25, f=f^*=2$ and $m=-1$. This configuration significantly favours hunting strategies over foraging strategies, whilst inducing an element of risk to hunting with a negative mauling reward. Furthermore, to align the gridworld stag hunt with a one-shot matrix game, we implemented an infinite-horizon setup. In this format, episodes conclude only when both agents have received a reward: through hunting, foraging, or being mauled. In asynchronous episode terminations, an agent that has received a reward is frozen on the grid, while the other agent continues to act. 

\subsubsection{Variable Complexity}
In Table~\ref{tab:gamedynamics} we introduce randomised and deterministic configurations for three environment parameters in the gridworld Stag Hunt: the spawn locations of agents, the spawn locations of the stag, and movement of the stag, where agent and stag spawn parameters apply to the start of every episode, and the stag movement applies to every timestep. Enumerating all combinations of values for these parameters provides eight unique environment variants, each with a specified degree of randomisation. These environment variants are outlined in Table \ref{tab:fullenvlist} and we use the corresponding environment label to denote the degree of randomisation in a given environment, i.e., the label \textbf{FFR} denotes an environment with: deterministic agent spawn (\textbf{F}), deterministic stag spawn (\textbf{F}) and randomised stag movement (\textbf{R}). We characterise the spectrum of randomisation across these eight variants as environment complexity in our study, where environments with randomisation on multiple parameters, such as environment \textbf{RRR}, are considered more complex than those with deterministic dynamics, such as environment \textbf{FFF}. 

\newcolumntype{P}[1]{>{\raggedright\arraybackslash}p{#1}}

\begin{table}[h]
\caption{Additional configurations for three environment parameters introduced to control complexity in the gridworld Stag Hunt.}
\centering
\begin{tabular}{llP{8cm}}
    \toprule
    \textbf{Environment Parameter} & \textbf{Value} & \textbf{Description} \\
    \midrule
    \multirow{2}{*}{Agent Spawn} & Fixed  & Blue and red agents spawn in the upper left and right corners of the grid, respectively.\\
    & Random & Blue and red agents spawn at random, distinct cells on the grid.\\
    \multirow{2}{*}{Stag Spawn} & Fixed & Stag spawns at the center cell of the grid.\\
    & Random & Stag spawns at a random cell location, distinct from either agent.\\
    \multirow{2}{*}{Stag Movement} & Follows & Stag moves to a neighbouring cell closest to the nearest agent, measured by Euclidean distance.\\
    & Random & Stag moves to a random neighbouring cell.\\
    \bottomrule
\end{tabular}
\label{tab:gamedynamics}
\end{table}

\begin{table}[h]
    \caption{The eight environment variants we create with our modifications to the gridworld Stag Hunt. Each row represents an environment with a specified degree of complexity.}
    \centering
    \begin{tabular}{llll}
        \toprule
        \textbf{Environment Label} & \textbf{Agent spawn} & \textbf{Stag spawn} & \textbf{Stag movement} \\
        \midrule
        \textbf{FFF} & Fixed & Fixed & Follows \\
        \textbf{RFF} & Random & Fixed & Follows \\
        \textbf{FRF} & Fixed & Random & Follows \\
        \textbf{FFR} & Fixed & Fixed & Random \\
        \textbf{RRF} & Random & Random & Follows \\
        \textbf{FRR} & Fixed & Random & Random \\
        \textbf{RFR} & Random & Fixed & Random \\
        \textbf{RRR} & Random & Random & Random \\
        \bottomrule
    \end{tabular} 
    \label{tab:fullenvlist}
\end{table}


\subsection{Experimental Setup}
Our experiments use RLLib's \citep{liang2018rllib} Proximal Policy Optimisation (PPO) \citep{schulman2017proximal} implementation to train pairs of agents in the gridworld Stag Hunt environment. We consider the simplest multi-agent variant of PPO, Independent PPO (IPPO), in which each agent uses an independent instance of PPO to optimise its own policy, through maximising individual rewards \cite{10.5555/284860.284934}. In particular, no parameters are shared between the multiple policies and each agent treats other agents as part of their environment observation. Note that, given the global observation space provided to agents in the gridworld Stag Hunt, our IPPO approach aligns closely with the widely used multi-agent shared-critic variant of PPO, MAPPO \cite{Yu2021TheSE}. We also consider a single-agent approach to the multi-agent problem, using fully-centralised PPO to serve as baseline to compare the performance of IPPO. In this approach, the observations of both agents are combined into a single observation, to train a fully centralised joint-action policy that controls both agents and maximises combined rewards. 

\section{Results}
\label{sec:Results}
\subsection{Environment Complexity and Strategy Convergence}
\label{sec:exp1}
We conduct five training runs of IPPO and PPO in each of the eight environments listed in Table \ref{tab:fullenvlist}, training for 1,000 iterations, each consisting of 4,000 environment timesteps. In each trial we use a distinct random seed for determining environment randomisation dynamics where applicable. We defer the results of our baseline, PPO, and the hyperparameters used, to Appendices \ref{app:hparams} and \ref{sec:CPPOExps}. 

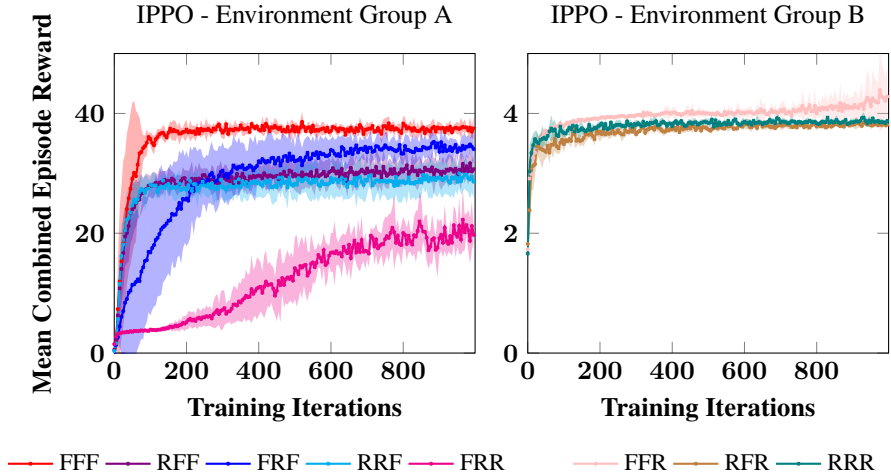
\begin{figure}[h]
    \hfill
        \begin{tikzpicture}
            \begin{groupplot}[
                group style={
                    group size=2 by 1, 
                    horizontal sep=0.7cm, 
                    group name=my plots, 
                    ylabels at=edge left,
                    xlabels at=edge bottom,
                    xticklabels at=edge bottom,
                },
                myplotstyle,
                ymin=0, 
                ymax=50,
                xmin=0,
                xmax=999,
                scale=0.7,
                every axis plot/.append style={
                font=\footnotesize,
                title style={font=\footnotesize},
                label style={font=\footnotesize},
                tick label style={font=\footnotesize},  
                legend style={font=\footnotesize}  
                } 
            ]
            \nextgroupplot[ylabel={Mean Combined Episode Reward}, yticklabel style={align=right},
            title={IPPO - Environment Group A},
            xlabel={Training Iterations},
            legend style={legend columns=-1, at={(0.4,-0.3)}, anchor=north}]

            \addplot+[color=red, smooth, name path=MAPPO_FFF_MEAN, mark=x, mark size=0.75, each nth point=5] table[x index=0, y index=6, col sep=comma] {experiment_1/MAPPO_FFF.csv}; 
            \addlegendentry{FFF}
            \addplot+[name path=MAPPO_FFF_UPPER, draw=none, opacity=0, forget plot, each nth point=5] table[x index=0, y expr=\thisrowno{6}+\thisrowno{7}, col sep=comma] {experiment_1/MAPPO_FFF.csv}; 
            \addplot+[name path=MAPPO_FFF_LOWER, draw=none, opacity=0, forget plot, each nth point=5] table[x index=0, y expr=\thisrowno{6}-\thisrowno{7}, col sep=comma] {experiment_1/MAPPO_FFF.csv}; 
            \addplot+[fill=red, fill opacity=0.3, forget plot, each nth point=5] fill between[of=MAPPO_FFF_UPPER and MAPPO_FFF_LOWER];

            \addplot+[color=violet, smooth, name path=MAPPO_RFF_MEAN, mark=x, mark size=0.75, each nth point=5 ] table[x index=0, y index=6, col sep=comma] {experiment_1/MAPPO_RFF.csv}; 
            \addlegendentry{RFF}
            \addplot+[name path=MAPPO_RFF_UPPER, draw=none, opacity=0, forget plot, each nth point=5] table[x index=0, y expr=\thisrowno{6}+\thisrowno{7}, col sep=comma] {experiment_1/MAPPO_RFF.csv};
            \addplot+[name path=MAPPO_RFF_LOWER, draw=none, opacity=0, forget plot, each nth point=5] table[x index=0, y expr=\thisrowno{6}-\thisrowno{7}, col sep=comma] {experiment_1/MAPPO_RFF.csv};
            \addplot+[fill=violet, fill opacity=0.3, forget plot, each nth point=5] fill between[of=MAPPO_RFF_UPPER and MAPPO_RFF_LOWER];

            \addplot+[color=blue, smooth, name path=MAPPO_FRF_MEAN, mark=x, mark size=0.75, each nth point=5 ] table[x index=0, y index=6, col sep=comma] {experiment_1/MAPPO_FRF.csv}; 
            \addlegendentry{FRF}
            \addplot+[name path=MAPPO_FRF_UPPER, draw=none, opacity=0, forget plot, each nth point=5] table[x index=0, y expr=\thisrowno{6}+\thisrowno{7}, col sep=comma] {experiment_1/MAPPO_FRF.csv};
            \addplot+[name path=MAPPO_FRF_LOWER, draw=none, opacity=0, forget plot, each nth point=5] table[x index=0, y expr=\thisrowno{6}-\thisrowno{7}, col sep=comma] {experiment_1/MAPPO_FRF.csv};
            \addplot+[fill=blue, fill opacity=0.3, forget plot, each nth point=5] fill between[of=MAPPO_FRF_UPPER and MAPPO_FRF_LOWER];

            \addplot+[color=cyan, smooth, name path=MAPPO_RRF_MEAN, mark=x, mark size=0.75, each nth point=5 ] table[x index=0, y index=6, col sep=comma] {experiment_1/MAPPO_RRF.csv}; 
            \addlegendentry{RRF}
            \addplot+[name path=MAPPO_RRF_UPPER, draw=none, opacity=0, forget plot, each nth point=5] table[x index=0, y expr=\thisrowno{6}+\thisrowno{7}, col sep=comma] {experiment_1/MAPPO_RRF.csv};
            \addplot+[name path=MAPPO_RRF_LOWER, draw=none, opacity=0, forget plot, each nth point=5] table[x index=0, y expr=\thisrowno{6}-\thisrowno{7}, col sep=comma] {experiment_1/MAPPO_RRF.csv};
            \addplot+[fill=cyan, fill opacity=0.3, forget plot, each nth point=5] fill between[of=MAPPO_RRF_UPPER and MAPPO_RRF_LOWER];

            \addplot+[color=magenta, smooth, name path=MAPPO_FRR_MEAN, mark=x, mark size=0.75, each nth point=5 ] table[x index=0, y index=6, col sep=comma] {experiment_1/MAPPO_FRR.csv}; 
            \addlegendentry{FRR}
            \addplot+[name path=MAPPO_FRR_UPPER, draw=none, opacity=0, forget plot, each nth point=5] table[x index=0, y expr=\thisrowno{6}+\thisrowno{7}, col sep=comma] {experiment_1/MAPPO_FRR.csv};
            \addplot+[name path=MAPPO_FRR_LOWER, draw=none, opacity=0, forget plot, each nth point=5] table[x index=0, y expr=\thisrowno{6}-\thisrowno{7}, col sep=comma] {experiment_1/MAPPO_FRR.csv};
            \addplot+[fill=magenta, fill opacity=0.3, forget plot, each nth point=5] fill between[of=MAPPO_FRR_UPPER and MAPPO_FRR_LOWER];

            \nextgroupplot[title={IPPO - Environment Group B}, ymin=0, ymax=5,
            xlabel={Training Iterations},
            legend style={legend columns=-1, at={(0.55,-0.3)}, anchor=north}]
            \addplot+[color=pink, smooth, name path=MAPPO_FFR_MEAN, mark=x, mark size=0.75, each nth point=5 ] table[x index=0, y index=6, col sep=comma] {experiment_1/MAPPO_FFR.csv}; 
            \addlegendentry{FFR}
            \addplot+[name path=MAPPO_FFR_UPPER, draw=none, opacity=0, forget plot, each nth point=5] table[x index=0, y expr=\thisrowno{6}+\thisrowno{7}, col sep=comma] {experiment_1/MAPPO_FFR.csv};
            \addplot+[name path=MAPPO_FFR_LOWER, draw=none, opacity=0, forget plot, each nth point=5] table[x index=0, y expr=\thisrowno{6}-\thisrowno{7}, col sep=comma] {experiment_1/MAPPO_FFR.csv};
            \addplot+[fill=pink, fill opacity=0.3, forget plot, each nth point=5] fill between[of=MAPPO_FFR_UPPER and MAPPO_FFR_LOWER];

            \addplot+[color=brown, smooth, name path=MAPPO_RFR_MEAN, mark=x, mark size=0.75, each nth point=5 ] table[x index=0, y index=6, col sep=comma] {experiment_1/MAPPO_RFR.csv}; 
            \addlegendentry{RFR}
            \addplot+[name path=MAPPO_RFR_UPPER, draw=none, opacity=0, forget plot, each nth point=5] table[x index=0, y expr=\thisrowno{6}+\thisrowno{7}, col sep=comma] {experiment_1/MAPPO_RFR.csv};
            \addplot+[name path=MAPPO_RFR_LOWER, draw=none, opacity=0, forget plot, each nth point=5] table[x index=0, y expr=\thisrowno{6}-\thisrowno{7}, col sep=comma] {experiment_1/MAPPO_RFR.csv};
            \addplot+[fill=brown, fill opacity=0.3, forget plot, each nth point=5] fill between[of=MAPPO_RFR_UPPER and MAPPO_RFR_LOWER];

            \addplot+[color=teal, smooth, name path=MAPPO_RRR_MEAN, mark=x, mark size=0.75, each nth point=5 ] table[x index=0, y index=6, col sep=comma] {experiment_1/MAPPO_RRR.csv}; 
            \addlegendentry{RRR}
            \addplot+[name path=MAPPO_RRR_UPPER, draw=none, opacity=0, forget plot, each nth point=5] table[x index=0, y expr=\thisrowno{6}+\thisrowno{7}, col sep=comma] {experiment_1/MAPPO_RRR.csv};
            \addplot+[name path=MAPPO_RRR_LOWER, draw=none, opacity=0, forget plot, each nth point=5] table[x index=0, y expr=\thisrowno{6}-\thisrowno{7}, col sep=comma] {experiment_1/MAPPO_RRR.csv};
            \addplot+[fill=teal, fill opacity=0.3, forget plot, each nth point=5] fill between[of=MAPPO_RRR_UPPER and MAPPO_RRR_LOWER];

        \end{groupplot}
    \end{tikzpicture}
    \hfill\null
    \caption{Training performance of IPPO in eight environments with varying degrees of complexity, detailed in Table \ref{tab:fullenvlist}. Environments are grouped by reward patterns: Group A (left) includes \textbf{FFF, RFF, FRF, RRF, FRR}, and Group B (right) includes \textbf{FFR, RFR, RRR}. Bold lines represent the average performance over five trials, and the shading represents ±1 standard deviation from each point.}
    \label{fig:4.1figure} 
\end{figure}

In Figure \ref{fig:4.1figure}, which shows the training performance of IPPO agents, measured by mean combined episode reward per iteration, averaged across 5 trials; we group environments based on observed reward patterns. Environments yielding high rewards (i.e., \textbf{FFF, RFF, FRF, RRF, FRR}) are denoted as Group A, while those with lower rewards (\textbf{FFR, RFR, RRR}) constitute Group B. These groupings reveal two insights into the learnt strategies of IPPO agents:

\paragraph{Observation 1: Agents trained in Group B converge to globally suboptimal strategies, consisting of bilateral foraging.} By analysing the mean episode rewards at the end of 1,000 training iterations, we infer agent strategies based on the previously established payoff structure. Specifically, exclusive bilateral hunting yields a combined reward of 50, significantly higher than the 4 from bilateral foraging. Consequently, stable convergence to a mean episode reward over 4 indicates that agents have adopted strategies that include bilateral hunting. Thus, in Group B, where the mean combined rewards rarely exceed 4 (second column of Figure \ref{fig:4.1figure}), we can conclude that agents have almost-exclusively converged to bilateral foraging strategies. 

\paragraph{Observation 2: Agents trained in Group A converge to mixed-strategies, consisting of both bilateral foraging and hunting.}  In Group A, where mean combined rewards consistently exceed the foraging threshold (1st column of Figure \ref{fig:4.1figure}), there is evident near-convergence to bilateral hunting strategies. However, we observe that agents in these environments do not achieve a consistent reward of 50, which is required to indicate convergence to exclusive bilateral hunting strategies. Specifically, in the 1,000th training iteration

\subsection{Escaping Suboptimal Strategies}
\label{sec:exp2}
We consider only Group B environments from the previous subsection, in which we observed IPPO agents converging to suboptimal strategies (Observation 1). To assess the feasibility of higher-reward strategies in these environments, we bias agents toward bilateral hunting strategies through a simple curriculum. We introduce a new environment variant, \textbf{cXXX}, designed to enforce a purely cooperative Nash equilibrium by eliminating the reward for foraging, thus exclusively incentivising bilateral hunting. This corresponds to a Stag Hunt game, as formalised in Table~\ref{tab:payoffMatrix}, where $h=25$, $f=f^{*}=0$ and $m=-1$. 

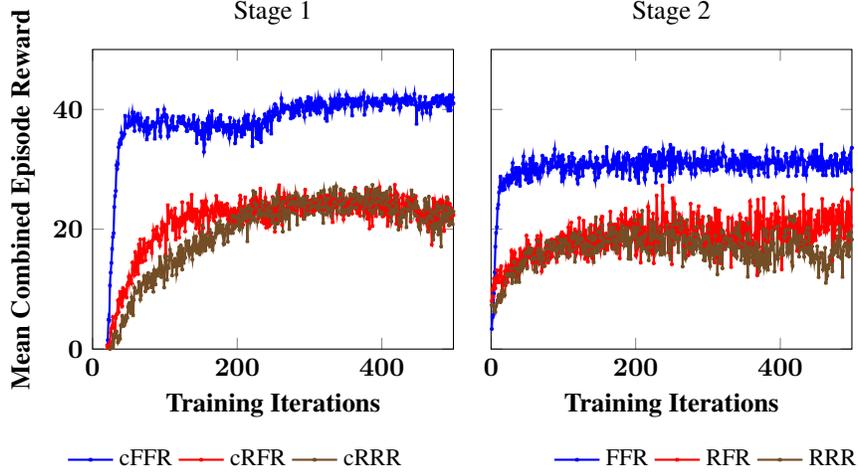
\begin{figure}[htb]
    \centering
        \begin{tikzpicture}
            \begin{groupplot}[
                group style={
                    group size=2 by 1, 
                    horizontal sep=0.5cm, 
                    group name=my plots, 
                    ylabels at=edge left,
                    xlabels at=edge bottom,
                    yticklabels at=edge left,
                    xticklabels at=edge bottom,
                },
                myplotstyle,
                ymin=0, 
                ymax=50,
                xmin=0,
                xmax=499,
                scale=0.7,
                every axis plot/.append style={
                font=\footnotesize,
                title style={font=\footnotesize},
                label style={font=\footnotesize},
                tick label style={font=\footnotesize},  
                legend style={font=\footnotesize}  
                } 
            ]
            \nextgroupplot[ylabel={Mean Combined Episode Reward}, yticklabel style={align=right},
            title={Stage 1}, xlabel={Training Iterations}, legend entries={cFFR, cRFR, cRRR}, legend style={legend columns=-1, at={(0.4,-0.3)}, anchor=north}]
            \addplot+[smooth, mark=x, mark size=0.75 ] table[x index=0, y index=1, col sep=comma] {curic_exps/curic_exps_final.csv};
            \addplot+[smooth, mark=x, mark size=0.75 ] table[x index=0, y index=3, col sep=comma] {curic_exps/curic_exps_final.csv};
            \addplot+[smooth, mark=x, mark size=0.75 ] table[x index=0, y index=5, col sep=comma] {curic_exps/curic_exps_final.csv};

            \nextgroupplot[yticklabels={}, title={Stage 2}, xlabel={Training Iterations}, legend entries={FFR, RFR, RRR},
            legend style={legend columns=-1, at={(0.6,-0.3)}, anchor=north}]
            \addplot+[smooth, mark=x, mark size=0.75 ] table[x index=0, y index=2, col sep=comma] {curic_exps/curic_exps_final.csv};
            \addplot+[smooth, mark=x, mark size=0.75 ] table[x index=0, y index=4, col sep=comma] {curic_exps/curic_exps_final.csv};
            \addplot+[smooth, mark=x, mark size=0.75 ] table[x index=0, y index=6, col sep=comma] {curic_exps/curic_exps_final.csv};

        \end{groupplot}
    \end{tikzpicture}
    \caption{Training performance of IPPO in a 2-stage curriculum in Group B environments from Section \ref{sec:exp1}. Agents are initially trained in a cooperation-inducing environment (\textbf{cXXX}), before being trained in their target environments.}
    \label{fig:4.2curiculumresults} 
\end{figure}

For each environment in Group B, a corresponding \textbf{cXXX} variant is created: \textbf{cFFR, cRFR} and \textbf{cRRR}. These variants maintain their respective randomisation dynamics described in Table \ref{tab:fullenvlist}. Initially, agents are trained in their respective \textbf{cXXX} environments for 500 iterations to learn hunting strategies (Stage 1). To discourage passive strategies and further promote hunting, a timestep penalty of -0.5 is introduced in these environments. Following this step, the pre-trained policies are then trained in their original environments (\textbf{FFR, RFR, RRR}) for another 500 iterations, during which the timestep penalty is removed (Stage 2). We use only IPPO for these experiments.

\textbf{Observation 3: Curriculum learning allows agents to escape suboptimal strategies in Group B.} Our results in Figure \ref{fig:4.2curiculumresults} demonstrate that our curriculum enables agents to find strategies in Group B environments that yield significantly higher rewards than those observed previously. In particular, during the second training stage, performance begins to converge toward average rewards that substantially exceed those attainable through exclusive foraging. Consistent learning of higher-reward strategies demonstrates that agents have not only discovered, but also converged to strategies that include bilateral hunting—effectively escaping the suboptimal strategies that were previously observed. 


\subsection{Empirical Game-Theoretic Analysis}
\label{sec:EGTA}
Here we further characterise the Group B environments with an empirical game-theoretic analysis following the methodology of \citet{Leibo2017MultiagentRL}. Recall from Section \ref{sec:exp1} that agents trained in these environments converged to suboptimal foraging strategies (Observation 1). In the context of the Stag Hunt, foraging corresponds to the defect action within a matrix game social dilemma (MGSD). Consequently, we categorise the strategies that these policies implement as defective. Conversely, in Section \ref{sec:exp2}, we present a learning curriculum that shifts agents from these suboptimal strategies towards cooperative hunting-focused strategies. Thus we can train two distinct types of policy for each environment in Group B: cooperative policies, $\Pi^C$, and defective policies, $\Pi^D$. This allows us to construct a meta-game with cooperate and defect actions that involve sampling a policy from either category and determining payoffs based on the resulting pairwise average rewards. This meta-game represents an MGSD, induced by the policy sets, $\Pi^C$ and $\Pi^D$, embedded within the underlying Markov Game, if such a mapping exists. We compute the empirical payoffs of this meta-game as follows: 

For each environment, two pairs of policies, $(\pi_{1}^C, \pi_{1}^D)$ and $(\pi_{2}^C, \pi_{2}^D)$, are sampled, where $\pi_{i}^C \in \Pi^C$ and $\pi_{j}^D \in \Pi^D$. These policy pairs represent the actions of a row player and a column player in a two-player matrix game with cooperate or defect actions (i.e., an MGSD), respectively. Specifically, $\pi_{i}^C$ corresponds to the cooperate action for player $i$, and $\pi_{j}^D$ corresponds to the defect action for player $j$. The outcomes (payoffs) for each combination of actions are determined by playing the respective policies against one another for an episode, in the respective Stag Hunt environment. The rewards obtained in each episode are then averaged in the corresponding cell in the payoff matrix. This procedure is repeated for each environment until convergence of all cell values, yielding the set of induced empirical payoff matrices shown in Table \ref{tab:EGTA}.

\begin{center}
    \begin{table}[ht]
    \centering
    \caption{Empirical payoff matrices for the meta-game in Group B environments. Each cell in the table contains a reward tuple: $(x,y)$, where $x$ is the reward for player/policy $1$, and $y$ is the reward for player/policy $2$, averaged over 5,000 episodes (2 d.p.)}
    \renewcommand{\arraystretch}{1.2} 
    \setlength{\tabcolsep}{4pt} 
    \begin{tabular}{@{}ccccccc@{}}
        \toprule
        & \multicolumn{2}{c}{RFR} & \multicolumn{2}{c}{RRR} & \multicolumn{2}{c}{FFR} \\
        \cmidrule(lr){2-3} \cmidrule(lr){4-5} \cmidrule(lr){6-7}
        & $\pi_{2}^C$& $\pi_{2}^D$& $\pi_{2}^C$& $\pi_{2}^D$& $\pi_{2}^C$& $\pi_{2}^D$\\
        \midrule
        $\pi_{1}^C$& (10.76, 10.81) & (0.45, 1.89) & (8.38, 8.37) & (0.42, 1.87) & (15.29, 15.26) & (0.49, 1.88) \\
        $\pi_{1}^D$& (1.95, 0.59) & \textbf{(1.93, 1.89)} & (1.96, 0.33) & \textbf{(1.95, 1.87)} & (2.02, 0.24) & \textbf{(1.98, 1.99)} \\
        \bottomrule
    \end{tabular}
    \label{tab:EGTA}
\end{table}
\end{center}

\paragraph{Observation 4: Group B environments map to MGSDs with Stag Hunt payoffs, and consequently, can be modelled as SSDs.} Section~\ref{sec:EGTA} maps a Markov Game to an MGSD, induced by a set of policies. To verify if this mapping is valid, here we examine the empirical payoffs of the constructed MGSD. It is important to note that it is not necessary for this process to generate empirical payoffs that constitute a valid MGSD, as the space of Markov Games is much larger than that of MGSDs. This is particularly relevant in the gridworld Stag Hunt we consider in this work, alongside our modifications, which has not been formally characterised in relation to the MGSD Stag Hunt. Nonetheless, the empirical payoff matrices in Table \ref{sec:EGTA} satisfy the five the MGSD inequalities in Equation \ref{eq:MGSDineq}, as well as the Stag Hunt inequality in Table \ref{tab:payoffMatrix}. The Markov Games that model Group B environments can be mapped to MGSD Stag Hunt games using the method described above. Consequently, they can indeed be modelled as Sequential Social Dilemmas (SSDs) (Section \ref{sec:markovGames}). 

\paragraph{Observation 5: Agents trained in Group B empirically converge to the suboptimal, risk-dominant Nash equilibrium strategy from the MGSD Stag Hunt.} Although formally verifying Nash equilibria in general-sum Markov Game policies can often be intractable \cite{pmlr-v54-perolat17a, 2022arXiv220802204K}, our analysis provides empirical evidence that agents in Section \ref{sec:exp1} have converged not only to suboptimal strategies (Insight 1), but also to Nash equilibria strategies. Consider the outcomes of two defecting policies, $(\pi_{1}^D, \pi_{2}^D)$, in Table \ref{tab:EGTA}. If agents had not converged to the suboptimal Nash equilibrium, we would see empirical payoffs for these outcomes exceeding a reward of 2 for each defecting policy. Instead, we observe convergence to a reward less than 2, indicating that when these policies attempt to deviate from the suboptimal strategy, they only reduce their own payoffs. This behaviour aligns precisely with the definition of Nash equilibria. Moreover, the specific equilibrium they converge to is reflective of the risk-dominant Nash equilibrium in the MGSD Stag Hunt (Section \ref{sec:MGSD}). 

\section{Discussion}
Our results indicate that agents trained in Group B environments converge to suboptimal strategies (Observation 1) and that these policies empirically represent Nash equilibria strategies (Observation 5). However, we also discover that agents are in-fact capable of learning more-optimal strategies within these environments (Observation 3). Despite the viability of higher-reward strategies, we observe that agents \textit{systematically} converge to the suboptimal Nash equilibrium strategy (i.e., mutual foraging). We suggest that this behaviour is a result of the increased complexity introduced by our modifications to the gridworld Stag Hunt. To support this, we explore three alternative explanations for our empirical findings:

First, although the observation space for each agent includes the positions of all entities within the grid, converging on higher-reward strategies in more complex environments, such as those in Group B, may require previous observations, actions, or rewards to be included in the agent's observation. However, this information is not available in a one-shot version of the game and thus cannot be considered a viable solution. Second, it could be that performing higher-reward strategies in Group B environments is significantly improbable due to highly randomised environment dynamics, and hence agents are behaving as expected, converging to the most optimal strategy (i.e., maximising long-term reward). We mitigate this proposition (and also the first explanation) by demonstrating empirically that through curriculum learning the same agents are able to learn significantly more cooperative strategies (Observation 3). Third, the dynamics of Group B environments might simply require more suitable hyperparameter values or additional training time, e.g., to increase exploration behaviour or allow more time to discover better strategies. Thus, in Appendix \ref{sec:exp1_extended}, we run further training experiments in Group B, including multiple hyperparameter searches, for 12,000 iterations. The results show that no further improvements are made under any circumstances, indicating that the observed behaviour is likely not an artifact of hyperparameter selection or insufficient training time, but a systematic defection toward lower-reward strategies due to increased environment complexity.

Thus, our findings suggest that the observed suboptimal convergence behaviour is a direct result of the increased complexity found in the Group B environments. Examining the environment dynamics within Group B indicates that the complexity introduced by a randomly moving stag—environments labelled by \textbf{xxR}—induces significant uncertainty which constrains learned strategies to suboptimal equilibria. Indeed, in three of the four environments featuring a randomly moving stag, our agents are consistently unable to learn cooperative strategies. Furthermore, environments with intermediate levels of complexity— i.e., those in-between \textbf{FFF} and \textbf{RRR}—maintain average rewards above the suboptimal strategy threshold (Observation 2) and show rewards decreasing proportionally to increased environment complexity. 

Finally, our results show that even within the framework of SSDs, multiple variants of the gridworld Stag Hunt can map to the same MGSD (Observation 4) and yet slight variations in environment dynamics, i.e., increased complexity, can promote convergence to vastly different strategies. Future work will seek to formalise Nash equilibria strategies in SSDs, and the notion of environment complexity, in the context of these empirical findings.


\section*{Acknowledgements}
Research was sponsored by the Army and was accomplished under Cooperative Agreement Number W911NF-22-2-0162. The views and conclusions contained in this document are those of the authors and should not be interpreted as representing the official policies, either expressed or implied, of the U.S. Army or the U.S. Government. The U.S. Government is authorized to reproduce and distribute reprints for Government purposes
notwithstanding any copyright notation herein. 

\bibliographystyle{plainnat}
\bibliography{references}

\begin{thebibliography}{35}
\providecommand{\natexlab}[1]{#1}
\providecommand{\url}[1]{\texttt{#1}}
\expandafter\ifx\csname urlstyle\endcsname\relax
  \providecommand{\doi}[1]{doi: #1}\else
  \providecommand{\doi}{doi: \begingroup \urlstyle{rm}\Url}\fi

\bibitem[Baker and Rachlin(2002)]{Baker2002Teaching}
Forest Baker and H.~Rachlin.
\newblock Teaching and learning in a probabilistic prisoner's dilemma.
\newblock \emph{Behavioural Processes}, 2002.

\bibitem[Bergstra et~al.(2013)Bergstra, Yamins, and Cox]{bergstra2013making}
James Bergstra, Daniel Yamins, and David Cox.
\newblock Making a science of model search: Hyperparameter optimization in hundreds of dimensions for vision architectures.
\newblock In \emph{International Conference on Machine Learning}, pages 115--123. PMLR, 2013.

\bibitem[Brockman et~al.(2016)Brockman, Cheung, Pettersson, Schneider, Schulman, Tang, and Zaremba]{1606.01540}
Greg Brockman, Vicki Cheung, Ludwig Pettersson, Jonas Schneider, John Schulman, Jie Tang, and Wojciech Zaremba.
\newblock Openai gym, 2016.

\bibitem[Carroll et~al.(2019)Carroll, Shah, Ho, Griffiths, Seshia, Abbeel, and Dragan]{10.5555/3454287.3454752}
Micah Carroll, Rohin Shah, Mark~K. Ho, Thomas~L. Griffiths, Sanjit~A. Seshia, Pieter Abbeel, and Anca Dragan.
\newblock \emph{On the utility of learning about humans for human-AI coordination}.
\newblock Curran Associates Inc., Red Hook, NY, USA, 2019.

\bibitem[Christianos et~al.(2020)Christianos, Sch{\"a}fer, and Albrecht]{christianos2020shared}
Filippos Christianos, Lukas Sch{\"a}fer, and Stefano Albrecht.
\newblock Shared experience actor-critic for multi-agent reinforcement learning.
\newblock \emph{Advances in neural information processing systems}, 33:\penalty0 10707--10717, 2020.

\bibitem[Dafoe et~al.(2020)Dafoe, Hughes, Bachrach, Collins, McKee, Leibo, Larson, and Graepel]{dafoe2020open}
Allan Dafoe, Edward Hughes, Yoram Bachrach, Tantum Collins, Kevin~R. McKee, Joel~Z. Leibo, Kate Larson, and Thore Graepel.
\newblock Open problems in cooperative ai, 2020.

\bibitem[Du et~al.(2023)Du, Leibo, Islam, Willis, and Sunehag]{Du_Leibo_Islam_Willis_Sunehag_2023}
Yali Du, Joel~Z. Leibo, Usman Islam, Richard Willis, and Peter Sunehag.
\newblock A review of cooperation in multi-agent learning.
\newblock \penalty0 (arXiv:2312.05162), December 2023.
\newblock \doi{10.48550/arXiv.2312.05162}.
\newblock arXiv:2312.05162 [cs].

\bibitem[Foerster et~al.(2018)Foerster, Chen, Al-Shedivat, Whiteson, Abbeel, and Mordatch]{10.5555/3237383.3237408}
Jakob Foerster, Richard~Y. Chen, Maruan Al-Shedivat, Shimon Whiteson, Pieter Abbeel, and Igor Mordatch.
\newblock Learning with opponent-learning awareness.
\newblock In \emph{Proceedings of the 17th International Conference on Autonomous Agents and MultiAgent Systems}, AAMAS '18, page 122–130, Richland, SC, 2018. International Foundation for Autonomous Agents and Multiagent Systems.

\bibitem[Jaderberg et~al.(2019)Jaderberg, Czarnecki, Dunning, Marris, Lever, Castañeda, Beattie, Rabinowitz, Morcos, Ruderman, Sonnerat, Green, Deason, Leibo, Silver, Hassabis, Kavukcuoglu, and Graepel]{Jaderberg_2019}
Max Jaderberg, Wojciech~M. Czarnecki, Iain Dunning, Luke Marris, Guy Lever, Antonio~Garcia Castañeda, Charles Beattie, Neil~C. Rabinowitz, Ari~S. Morcos, Avraham Ruderman, Nicolas Sonnerat, Tim Green, Louise Deason, Joel~Z. Leibo, David Silver, Demis Hassabis, Koray Kavukcuoglu, and Thore Graepel.
\newblock Human-level performance in 3d multiplayer games with population-based reinforcement learning.
\newblock \emph{Science}, 364\penalty0 (6443):\penalty0 859–865, May 2019.
\newblock ISSN 1095-9203.
\newblock \doi{10.1126/science.aau6249}.

\bibitem[{Kalogiannis} et~al.(2022){Kalogiannis}, {Anagnostides}, {Panageas}, {Vlatakis-Gkaragkounis}, {Chatziafratis}, and {Stavroulakis}]{2022arXiv220802204K}
Fivos {Kalogiannis}, Ioannis {Anagnostides}, Ioannis {Panageas}, Emmanouil-Vasileios {Vlatakis-Gkaragkounis}, Vaggos {Chatziafratis}, and Stelios {Stavroulakis}.
\newblock {Efficiently Computing Nash Equilibria in Adversarial Team Markov Games}.
\newblock \emph{arXiv e-prints}, art. arXiv:2208.02204, August 2022.
\newblock \doi{10.48550/arXiv.2208.02204}.

\bibitem[Khan et~al.(2024)Khan, Willi, Kwan, Tacchetti, Lu, Grefenstette, Rockt\"{a}schel, and Foerster]{10.5555/3635637.3662955}
Akbir Khan, Timon Willi, Newton Kwan, Andrea Tacchetti, Chris Lu, Edward Grefenstette, Tim Rockt\"{a}schel, and Jakob Foerster.
\newblock Scaling opponent shaping to high dimensional games.
\newblock In \emph{Proceedings of the 23rd International Conference on Autonomous Agents and Multiagent Systems}, AAMAS '24, page 1001–1010, Richland, SC, 2024. International Foundation for Autonomous Agents and Multiagent Systems.
\newblock ISBN 9798400704864.

\bibitem[Leibo et~al.(2017)Leibo, Zambaldi, Lanctot, Marecki, and Graepel]{Leibo2017MultiagentRL}
Joel~Z. Leibo, Vin{\'i}cius~Flores Zambaldi, Marc Lanctot, Janusz Marecki, and Thore Graepel.
\newblock Multi-agent reinforcement learning in sequential social dilemmas.
\newblock In \emph{Adaptive Agents and Multi-Agent Systems}, 2017.

\bibitem[Leibo et~al.(2021)Leibo, Due{\~n}ez-Guzman, Vezhnevets, Agapiou, Sunehag, Koster, Matyas, Beattie, Mordatch, and Graepel]{leibo2021scalable}
Joel~Z Leibo, Edgar~A Due{\~n}ez-Guzman, Alexander Vezhnevets, John~P Agapiou, Peter Sunehag, Raphael Koster, Jayd Matyas, Charlie Beattie, Igor Mordatch, and Thore Graepel.
\newblock Scalable evaluation of multi-agent reinforcement learning with melting pot.
\newblock In \emph{International conference on machine learning}, pages 6187--6199. PMLR, 2021.

\bibitem[Lerer and Peysakhovich(2018)]{lerer2018maintaining}
Adam Lerer and Alexander Peysakhovich.
\newblock Maintaining cooperation in complex social dilemmas using deep reinforcement learning, 2018.

\bibitem[Li et~al.(2020)Li, Jamieson, Rostamizadeh, Gonina, Ben-Tzur, Hardt, Recht, and Talwalkar]{li2020system}
Liam Li, Kevin Jamieson, Afshin Rostamizadeh, Ekaterina Gonina, Jonathan Ben-Tzur, Moritz Hardt, Benjamin Recht, and Ameet Talwalkar.
\newblock A system for massively parallel hyperparameter tuning.
\newblock \emph{Proceedings of Machine Learning and Systems}, 2:\penalty0 230--246, 2020.

\bibitem[Liang et~al.(2018)Liang, Liaw, Nishihara, Moritz, Fox, Goldberg, Gonzalez, Jordan, and Stoica]{liang2018rllib}
Eric Liang, Richard Liaw, Robert Nishihara, Philipp Moritz, Roy Fox, Ken Goldberg, Joseph Gonzalez, Michael Jordan, and Ion Stoica.
\newblock Rllib: Abstractions for distributed reinforcement learning.
\newblock In \emph{International conference on machine learning}, pages 3053--3062. PMLR, 2018.

\bibitem[Littman(1994)]{10.5555/3091574.3091594}
Michael~L. Littman.
\newblock Markov games as a framework for multi-agent reinforcement learning.
\newblock In \emph{Proceedings of the Eleventh International Conference on International Conference on Machine Learning}, ICML'94, page 157–163, San Francisco, CA, USA, 1994. Morgan Kaufmann Publishers Inc.
\newblock ISBN 1558603352.

\bibitem[Nash(1950)]{Nash_1950}
John~F. Nash.
\newblock Equilibrium points in n-person games.
\newblock \emph{Proceedings of the National Academy of Sciences}, 36\penalty0 (1):\penalty0 48–49, 1950.
\newblock \doi{10.1073/pnas.36.1.48}.

\bibitem[Nesterov-Rappoport(2022)]{nesterov-rappoport_2022}
David~Lvovich Nesterov-Rappoport.
\newblock The evolution of trust: Understanding prosocial behavior in multi-agent reinforcement learning systems.
\newblock Technical report, 2022.

\bibitem[Nica et~al.(2017)Nica, Berariu, Gogianu, and Florea]{8531285}
Andrei~Cristian Nica, Tudor Berariu, Florin Gogianu, and Adina~Magda Florea.
\newblock Learning to maximize return in a stag hunt collaborative scenario through deep reinforcement learning.
\newblock In \emph{2017 19th International Symposium on Symbolic and Numeric Algorithms for Scientific Computing (SYNASC)}, pages 188--195, 2017.
\newblock \doi{10.1109/SYNASC.2017.00039}.

\bibitem[Perolat et~al.(2017)Perolat, Strub, Piot, and Pietquin]{pmlr-v54-perolat17a}
Julien Perolat, Florian Strub, Bilal Piot, and Olivier Pietquin.
\newblock {Learning Nash Equilibrium for General-Sum Markov Games from Batch Data}.
\newblock In Aarti Singh and Jerry Zhu, editors, \emph{Proceedings of the 20th International Conference on Artificial Intelligence and Statistics}, volume~54 of \emph{Proceedings of Machine Learning Research}, pages 232--241. PMLR, 20--22 Apr 2017.

\bibitem[Peysakhovich and Lerer(2017)]{Peysakhovich2017ProsocialLA}
Alexander Peysakhovich and Adam Lerer.
\newblock Prosocial learning agents solve generalized stag hunts better than selfish ones.
\newblock In \emph{Adaptive Agents and Multi-Agent Systems}, 2017.

\bibitem[Schelling(1960)]{Schelling1960}
T.C. Schelling.
\newblock \emph{{The Strategy of Conflict}}.
\newblock Harvard University Press, 1960.

\bibitem[Schulman et~al.(2017)Schulman, Wolski, Dhariwal, Radford, and Klimov]{schulman2017proximal}
John Schulman, Filip Wolski, Prafulla Dhariwal, Alec Radford, and Oleg Klimov.
\newblock Proximal policy optimization algorithms, 2017.

\bibitem[Shapley(1953)]{Shapley_1953}
L.~S. Shapley.
\newblock Stochastic games*.
\newblock \emph{Proceedings of the National Academy of Sciences}, 39\penalty0 (10):\penalty0 1095–1100, 1953.
\newblock \doi{10.1073/pnas.39.10.1095}.

\bibitem[Smith(1976)]{Smith1976EvolutionAT}
John~Maynard Smith.
\newblock Evolution and the theory of games.
\newblock 1976.

\bibitem[Sunehag et~al.(2018)Sunehag, Lever, Gruslys, Czarnecki, Zambaldi, Jaderberg, Lanctot, Sonnerat, Leibo, Tuyls, and Graepel]{10.5555/3237383.3238080}
Peter Sunehag, Guy Lever, Audrunas Gruslys, Wojciech~Marian Czarnecki, Vinicius Zambaldi, Max Jaderberg, Marc Lanctot, Nicolas Sonnerat, Joel~Z. Leibo, Karl Tuyls, and Thore Graepel.
\newblock Value-decomposition networks for cooperative multi-agent learning based on team reward.
\newblock In \emph{Proceedings of the 17th International Conference on Autonomous Agents and MultiAgent Systems}, AAMAS '18, page 2085–2087. International Foundation for Autonomous Agents and Multiagent Systems, 2018.

\bibitem[Szolnoki and Perc(2009)]{Szolnoki_2009}
Attila Szolnoki and Matjaž Perc.
\newblock Resolving social dilemmas on evolving random networks.
\newblock \emph{EPL (Europhysics Letters)}, 2009.

\bibitem[Tan(1997)]{10.5555/284860.284934}
Ming Tan.
\newblock \emph{Multi-agent reinforcement learning: independent vs. cooperative agents}, page 487–494.
\newblock Morgan Kaufmann Publishers Inc., San Francisco, CA, USA, 1997.
\newblock ISBN 1558604952.

\bibitem[Terry et~al.(2021)Terry, Black, Grammel, Jayakumar, Hari, Sullivan, Santos, Dieffendahl, Horsch, Perez-Vicente, et~al.]{terry2021pettingzoo}
Jordan Terry, Benjamin Black, Nathaniel Grammel, Mario Jayakumar, Ananth Hari, Ryan Sullivan, Luis~S Santos, Clemens Dieffendahl, Caroline Horsch, Rodrigo Perez-Vicente, et~al.
\newblock Pettingzoo: Gym for multi-agent reinforcement learning.
\newblock \emph{Advances in Neural Information Processing Systems}, 34:\penalty0 15032--15043, 2021.

\bibitem[van Rooij(2007)]{Rooij2007TheSH}
Robert van Rooij.
\newblock The stag hunt and the evolution of social structure.
\newblock \emph{Studia Logica}, 85:\penalty0 133--138, 2007.

\bibitem[Vinyals et~al.(2019)Vinyals, Babuschkin, Czarnecki, Mathieu, Dudzik, Chung, Choi, Powell, Ewalds, Georgiev, Oh, Horgan, Kroiss, Danihelka, Huang, Sifre, Cai, Agapiou, Jaderberg, Vezhnevets, Leblond, Pohlen, Dalibard, Budden, Sulsky, Molloy, Paine, Gulcehre, Wang, Pfaff, Wu, Ring, Yogatama, W{\"u}nsch, McKinney, Smith, Schaul, Lillicrap, Kavukcuoglu, Hassabis, Apps, and Silver]{Vinyals2019GrandmasterLI}
Oriol Vinyals, Igor Babuschkin, Wojciech~M. Czarnecki, Micha{\"e}l Mathieu, Andrew Dudzik, Junyoung Chung, David Choi, Richard Powell, Timo Ewalds, Petko Georgiev, Junhyuk Oh, Dan Horgan, Manuel Kroiss, Ivo Danihelka, Aja Huang, L.~Sifre, Trevor Cai, John~P. Agapiou, Max Jaderberg, Alexander~Sasha Vezhnevets, R{\'e}mi Leblond, Tobias Pohlen, Valentin Dalibard, David Budden, Yury Sulsky, James Molloy, Tom~Le Paine, Caglar Gulcehre, Ziyun Wang, Tobias Pfaff, Yuhuai Wu, Roman Ring, Dani Yogatama, Dario W{\"u}nsch, Katrina McKinney, Oliver Smith, Tom Schaul, Timothy~P. Lillicrap, Koray Kavukcuoglu, Demis Hassabis, Chris Apps, and David Silver.
\newblock Grandmaster level in starcraft ii using multi-agent reinforcement learning.
\newblock \emph{Nature}, 575:\penalty0 350 -- 354, 2019.

\bibitem[Wellman(2006)]{Wellman2006MethodsFE}
Michael~P. Wellman.
\newblock Methods for empirical game-theoretic analysis.
\newblock In \emph{AAAI Conference on Artificial Intelligence}, 2006.

\bibitem[Yang et~al.(2007)Yang, Yue, and Yu]{Yang2007The}
Chun-Lei Yang, Ching-Syang~Jack Yue, and I.~Yu.
\newblock The rise of cooperation in correlated matching prisoners dilemma: An experiment.
\newblock \emph{Experimental Economics}, 2007.

\bibitem[Yu et~al.(2021)Yu, Velu, Vinitsky, Wang, Bayen, and Wu]{Yu2021TheSE}
Chao Yu, Akash Velu, Eugene Vinitsky, Yu~Wang, Alexandre~M. Bayen, and Yi~Wu.
\newblock The surprising effectiveness of ppo in cooperative multi-agent games.
\newblock In \emph{Neural Information Processing Systems}, 2021.

\end{thebibliography}
\medskip


\appendix
\section{Gridworld Stag Hunt}
\label{app:ENV}

\begin{figure}[h]
\caption{Rendering of the gridworld Stag Hunt environment used in this study.}
    \centering
    \includegraphics[height=5cm, width=5cm]{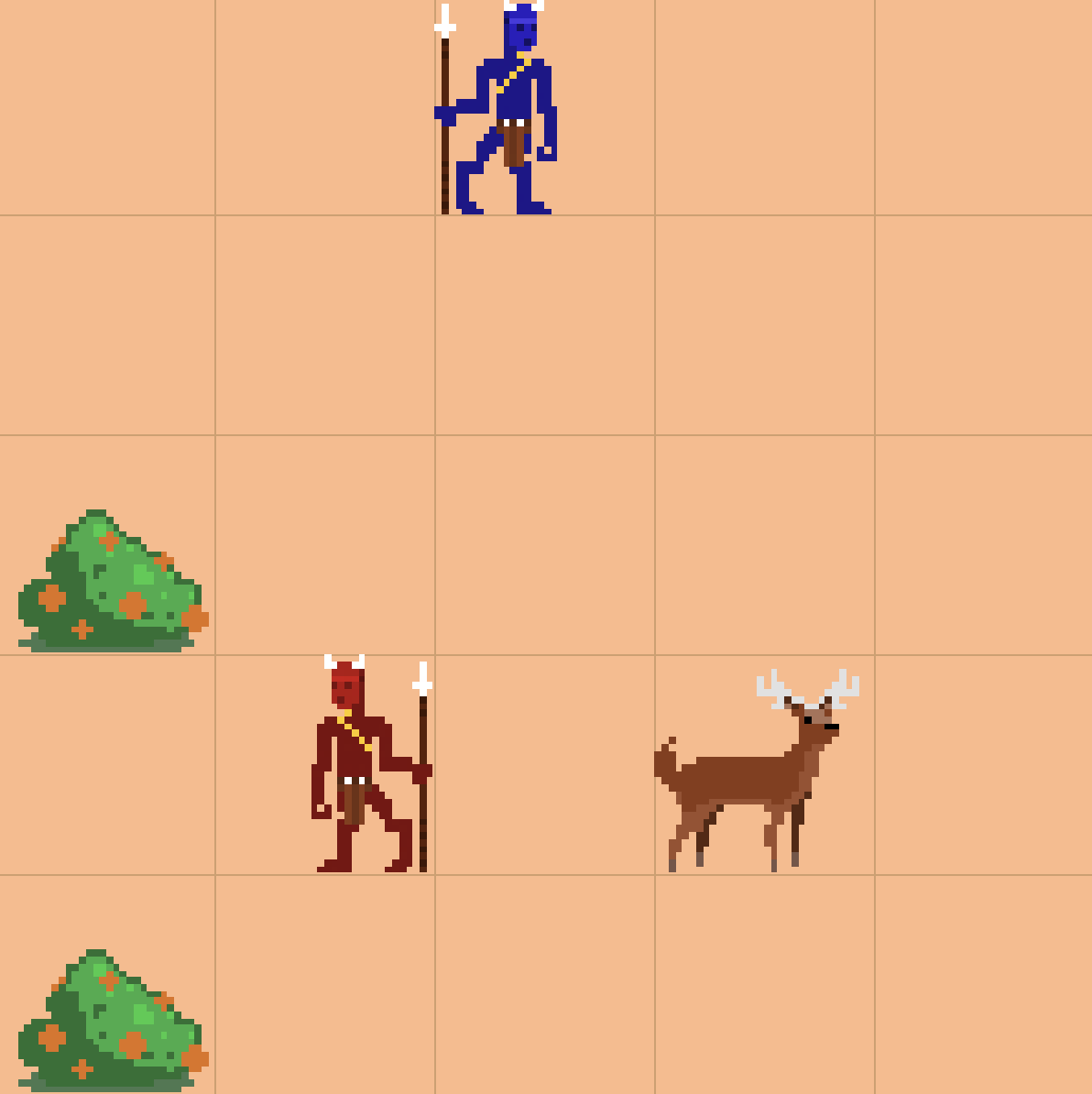}
    \label{fig:stagSS}
\end{figure}

\section{Hyperparameters}
\label{app:hparams}
In Table \ref{tab:hyperparameters}, we list the relevant hyperparameters for RLlib's PPO implementation which was employed throughout this study, using the stated hyperparameter values. All parameters not stated were set to None.
\begin{table}[ht]
\centering
\caption{Hyperparameter values used for training PPO policies throughout this study. The second column includes the range of values used for hyperparameter tuning in Section \ref{sec:exp1_extended}, where values given in square brackets [x,y,z] represent a discrete choice over values x, y and z, and values in round brackets (x,y) represent a continuous range, sampled according to the stated distribution. }
\label{tab:hyperparameters}
\begin{tabular}{@{}lcc@{}}
\toprule
Hyperparameter & IPPO/PPO (Section \ref{sec:Results}) & IPPO Extended (Section \ref{sec:exp1_extended}) \\ 
\midrule
clip & $0.3$ & $[0.2, 0.3, 0.4, 0.45]$\\
discount & $0.99$ & $[0.98, 0.99, 0.995, 0.999]$\\
training batch size & $4000$ & $4000$ \\
sgd batch size & $128$ & $[128, 256, 512]$\\
num\_sgd\_iter & $30$ & $30$ \\
kl coeffient & $0.2$ & $0.2$ \\
kl target & $0.2$ & $0.2$ \\
entropy coefficient & $0$ & $log\_uniform (5 \times 10^{-2}, 3 \times 10^{-1})$\\
vf clip & $10.0$ & $10.0$ \\
vf loss coefficient & $1.0$ & $uniform (0.5, 1.0)$ \\
$\lambda$ (GAE) & $1.0$ & $[0.92, 0.95, 0.98]$ \\
optimizer & Adam & Adam \\
learning rate& $5 \times 10^{-5}$& $log\_uniform (1 \times 10^{-6}, 5 \times 10^{-2})$\\
\bottomrule
\end{tabular}
\end{table}

\section{Environment Complexity and Strategy Convergence}
In this section, we provide results of the supplementary experiments we conducted to substantiate our claims in Section \ref{sec:exp1}.

\subsection{Centralised PPO}
\label{sec:CPPOExps}
To establish a baseline for our earlier results in Section \ref{sec:exp1}, which used the IPPO training methodology, we trained a single joint-action policy that controls both agents, labelled PPO. The distinction between IPPO and PPO is that while IPPO agents treat other learners as part of their own environment, PPO controls both agents under a single policy. This distinction allows us to assess the extent to which independent multi-agent interaction is necessary to achieve globally optimal outcomes in the gridworld Stag Hunt. 

\begin{figure}[h]
    \centering
        \begin{tikzpicture}
            \begin{groupplot}[
                group style={
                    group size=2 by 1, 
                    horizontal sep=0.7cm, 
                    group name=my plots, 
                    ylabels at=edge left,
                    xlabels at=edge bottom,
                    xticklabels at=edge bottom,
                },
                myplotstyle,
                ymin=0, 
                ymax=50,
                xmin=0,
                xmax=999,
                scale=0.7,
                every axis plot/.append style={
                font=\footnotesize,
                title style={font=\footnotesize},
                label style={font=\footnotesize},
                tick label style={font=\footnotesize},  
                legend style={font=\footnotesize}  
                } 
            ]
            
            \nextgroupplot[ylabel={Mean Combined Episode Reward}, yticklabel style={align=right},
            title={PPO - Environment Group A}, xlabel={Training Iterations},
            legend style={legend columns=-1, at={(0.4,-0.25)}, anchor=north}]

            \addplot+[color=red, smooth, name path=PPO_FFF_MEAN, mark=x, mark size=0.75, each nth point=5] table[x index=0, y index=6, col sep=comma] {experiment_1/PPO_FFF.csv}; 
            \addlegendentry{FFF}
            \addplot+[name path=PPO_FFF_UPPER, draw=none, opacity=0, forget plot, each nth point=5] table[x index=0, y expr=\thisrowno{6}+\thisrowno{7}, col sep=comma] {experiment_1/PPO_FFF.csv}; 
            \addplot+[name path=PPO_FFF_LOWER, draw=none, opacity=0, forget plot, each nth point=5] table[x index=0, y expr=\thisrowno{6}-\thisrowno{7}, col sep=comma] {experiment_1/PPO_FFF.csv}; 
            \addplot+[fill=red, fill opacity=0.3, forget plot, each nth point=5] fill between[of=PPO_FFF_UPPER and PPO_FFF_LOWER];

            \addplot+[color=violet, smooth, name path=PPO_RFF_MEAN, mark=x, mark size=0.75, each nth point=5 ] table[x index=0, y index=6, col sep=comma] {experiment_1/PPO_RFF.csv}; 
            \addlegendentry{RFF}
            \addplot+[name path=PPO_RFF_UPPER, draw=none, opacity=0, forget plot, each nth point=5] table[x index=0, y expr=\thisrowno{6}+\thisrowno{7}, col sep=comma] {experiment_1/PPO_RFF.csv};
            \addplot+[name path=PPO_RFF_LOWER, draw=none, opacity=0, forget plot, each nth point=5] table[x index=0, y expr=\thisrowno{6}-\thisrowno{7}, col sep=comma] {experiment_1/PPO_RFF.csv};
            \addplot+[fill=violet, fill opacity=0.3, forget plot, each nth point=5] fill between[of=PPO_RFF_UPPER and PPO_RFF_LOWER];

            \addplot+[color=blue, smooth, name path=PPO_FRF_MEAN, mark=x, mark size=0.75, each nth point=5 ] table[x index=0, y index=6, col sep=comma] {experiment_1/PPO_FRF.csv}; 
            \addlegendentry{FRF}
            \addplot+[name path=PPO_FRF_UPPER, draw=none, opacity=0, forget plot, each nth point=5] table[x index=0, y expr=\thisrowno{6}+\thisrowno{7}, col sep=comma] {experiment_1/PPO_FRF.csv};
            \addplot+[name path=PPO_FRF_LOWER, draw=none, opacity=0, forget plot, each nth point=5] table[x index=0, y expr=\thisrowno{6}-\thisrowno{7}, col sep=comma] {experiment_1/PPO_FRF.csv};
            \addplot+[fill=blue, fill opacity=0.3, forget plot, each nth point=5] fill between[of=PPO_FRF_UPPER and PPO_FRF_LOWER];

            \addplot+[color=cyan, smooth, name path=PPO_RRF_MEAN, mark=x, mark size=0.75, each nth point=5 ] table[x index=0, y index=6, col sep=comma] {experiment_1/PPO_RRF.csv}; 
            \addlegendentry{RRF}
            \addplot+[name path=PPO_RRF_UPPER, draw=none, opacity=0, forget plot, each nth point=5] table[x index=0, y expr=\thisrowno{6}+\thisrowno{7}, col sep=comma] {experiment_1/PPO_RRF.csv};
            \addplot+[name path=PPO_RRF_LOWER, draw=none, opacity=0, forget plot, each nth point=5] table[x index=0, y expr=\thisrowno{6}-\thisrowno{7}, col sep=comma] {experiment_1/PPO_RRF.csv};
            \addplot+[fill=cyan, fill opacity=0.3, forget plot, each nth point=5] fill between[of=PPO_RRF_UPPER and PPO_RRF_LOWER];

            \addplot+[color=magenta, smooth, name path=PPO_FRR_MEAN, mark=x, mark size=0.75, each nth point=5 ] table[x index=0, y index=6, col sep=comma] {experiment_1/PPO_FRR.csv}; 
            \addlegendentry{FRR}
            \addplot+[name path=PPO_FRR_UPPER, draw=none, opacity=0, forget plot, each nth point=5] table[x index=0, y expr=\thisrowno{6}+\thisrowno{7}, col sep=comma] {experiment_1/PPO_FRR.csv};
            \addplot+[name path=PPO_FRR_LOWER, draw=none, opacity=0, forget plot, each nth point=5] table[x index=0, y expr=\thisrowno{6}-\thisrowno{7}, col sep=comma] {experiment_1/PPO_FRR.csv};
            \addplot+[fill=magenta, fill opacity=0.3, forget plot, each nth point=5] fill between[of=PPO_FRR_UPPER and PPO_FRR_LOWER];

            \nextgroupplot[title={PPO - Environment Group B}, xlabel={Training Iterations},
            legend style={legend columns=-1, at={(0.5,-0.25)}, anchor=north},
            ymin=0, ymax=5]
            \addplot+[color=pink, smooth, name path=PPO_FFR_MEAN, mark=x, mark size=0.75, each nth point=5 ] table[x index=0, y index=6, col sep=comma] {experiment_1/PPO_FFR.csv}; 
            \addlegendentry{FFR}
            \addplot+[name path=PPO_FFR_UPPER, draw=none, opacity=0, forget plot, each nth point=5] table[x index=0, y expr=\thisrowno{6}+\thisrowno{7}, col sep=comma] {experiment_1/PPO_FFR.csv};
            \addplot+[name path=PPO_FFR_LOWER, draw=none, opacity=0, forget plot, each nth point=5] table[x index=0, y expr=\thisrowno{6}-\thisrowno{7}, col sep=comma] {experiment_1/PPO_FFR.csv};
            \addplot+[fill=pink, fill opacity=0.3, forget plot, each nth point=5] fill between[of=PPO_FFR_UPPER and PPO_FFR_LOWER];

            \addplot+[color=brown, smooth, name path=PPO_RFR_MEAN, mark=x, mark size=0.75, each nth point=5 ] table[x index=0, y index=6, col sep=comma] {experiment_1/PPO_RFR.csv}; 
            \addlegendentry{RFR}
            \addplot+[name path=PPO_RFR_UPPER, draw=none, opacity=0, forget plot, each nth point=5] table[x index=0, y expr=\thisrowno{6}+\thisrowno{7}, col sep=comma] {experiment_1/PPO_RFR.csv};
            \addplot+[name path=PPO_RFR_LOWER, draw=none, opacity=0, forget plot, each nth point=5] table[x index=0, y expr=\thisrowno{6}-\thisrowno{7}, col sep=comma] {experiment_1/PPO_RFR.csv};
            \addplot+[fill=brown, fill opacity=0.3, forget plot, each nth point=5] fill between[of=PPO_RFR_UPPER and PPO_RFR_LOWER];

            \addplot+[color=teal, smooth, name path=PPO_RRR_MEAN, mark=x, mark size=0.75, each nth point=5 ] table[x index=0, y index=6, col sep=comma] {experiment_1/PPO_RRR.csv}; 
            \addlegendentry{RRR}
            \addplot+[name path=PPO_RRR_UPPER, draw=none, opacity=0, forget plot, each nth point=5] table[x index=0, y expr=\thisrowno{6}+\thisrowno{7}, col sep=comma] {experiment_1/PPO_RRR.csv};
            \addplot+[name path=PPO_RRR_LOWER, draw=none, opacity=0, forget plot, each nth point=5] table[x index=0, y expr=\thisrowno{6}-\thisrowno{7}, col sep=comma] {experiment_1/PPO_RRR.csv};
            \addplot+[fill=teal, fill opacity=0.3, forget plot, each nth point=5] fill between[of=PPO_RRR_UPPER and PPO_RRR_LOWER];
        \end{groupplot}
    \end{tikzpicture}
    \caption{Training performance of PPO across eight environments of varying complexity, with labels detailed Table \ref{tab:fullenvlist}. Environments are grouped by reward patterns: Group A (left) includes \textbf{FFF, RFF, FRF, RRF, FRR}, and Group B (right) includes \textbf{FFR, RFR, RRR}. Bold lines represent the average performance across five trials and the shading represents ±1 standard deviation from each point.}
    \label{fig:PPO_exp1} 
\end{figure}
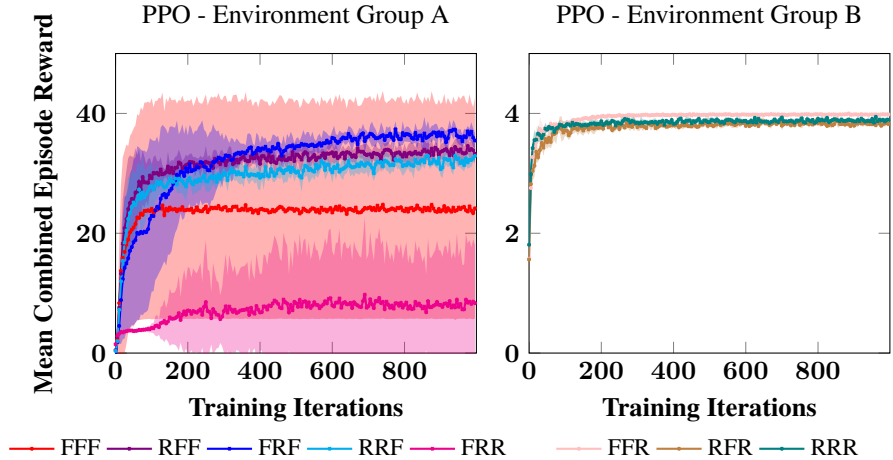

Our results, presented in Figure \ref{fig:PPO_exp1}, further suggest that environment complexity promotes convergence to suboptimal strategies. We find that the same grouping of environments from IPPO (Figure \ref{fig:4.1figure}) applies to PPO agents. Although the performance of PPO in Group B resembles that of IPPO in Group B, we observe that performance in Group A exhibits significantly larger variance between trials compared to IPPO. Specifically, multiple trials in the same environment, including the fully deterministic environment, \textbf{FFF}, show convergence to vastly different strategies. This observation highlights the necessity of independent multi-agent interaction in navigating the social dilemma aspects of our environment, where PPO may be overly influenced by early biases in training runs. Despite this, in Figure \ref{fig:PPO_exp1} we continue to observe that an increase in environment complexity correlates with decreased pay-offs. 

\subsection{Extended Training in Group B Environments}
\label{sec:exp1_extended}
In Section \ref{sec:exp1}, we found that IPPO agents converged to suboptimal strategies in Group B, whilst converging to more optimal strategies in Group A.  To investigate whether the observed suboptimal convergence behaviour could be explained through either: inappropriate hyperparameters for the given environment variants, or insufficient training time to discover more optimal strategies, we conducted an extended training experiment with hyperparameter tuning. 

Initially, we trained IPPO agents in each of the group B environments (\textbf{FFR, RFR, RRR}) for five trials, each consisting of 1,000 iterations. Each of these trials used HyperOpt's Tree Parzen Estimator (TSE) search algorithm \cite{bergstra2013making} and the Asynchronous Hyperband (ASHA) scheduler \cite{li2020system} to estimate the optimal hyperparameter values for each trial, where the hyperparameter ranges used are listed in the second column of Table \ref{tab:hyperparameters}. Subsequently, we selected the best performing trial in each environment, measured by maximum mean combined episode reward, to extract the optimal set of hyperparameter values for each environment. Finally, we used the set of optimal hyperparameter values to run an additional training experiment for each environment, consisting of 12,000 iterations, i.e., $4.8 \times 10^7$ sampled timesteps, representing a tenfold increase over the training duration used in Section \ref{sec:exp1}. 

\begin{figure}
    \centering
    \begin{tikzpicture}
        \begin{axis}[
            myplotstyle,  
            ymin=0, 
            ymax=5,
            xmin=0,
            xmax=11999,
            scale=1.0,
            title={IPPO Extended - Environment Group B},
            title style={align=center},
            xlabel={Training Iterations},
            ylabel={Mean Combined Episode Reward},
            tick label style={font=\footnotesize}, 
            legend style={at={(0.5,-0.1)}, anchor=north, legend columns = 3}, 
            legend entries={FFR, RFR, RRR},
            tick label style={font=\footnotesize},  
        ]
        \addplot+[color=pink, smooth, mark=x, mark size=1] table[x index=0, y index=3, col sep=comma] {experiment_2/longexps_filtered.csv}; 
        \addplot+[color=brown, smooth, mark=x, mark size=1] table[x index=0, y index=2, col sep=comma] {experiment_2/longexps_filtered.csv}; 
        \addplot+[color=teal, smooth, mark=x, mark size=1] table[x index=0, y index=1, col sep=comma] {experiment_2/longexps_filtered.csv}; 
        \end{axis}
    \end{tikzpicture}
    \caption{Training performance of IPPO in Group B environments from Section \ref{sec:exp1}, using the best hyperparameters found from five trials consisting of 1,000 iterations each.}
    \label{fig:longexps}
\end{figure}
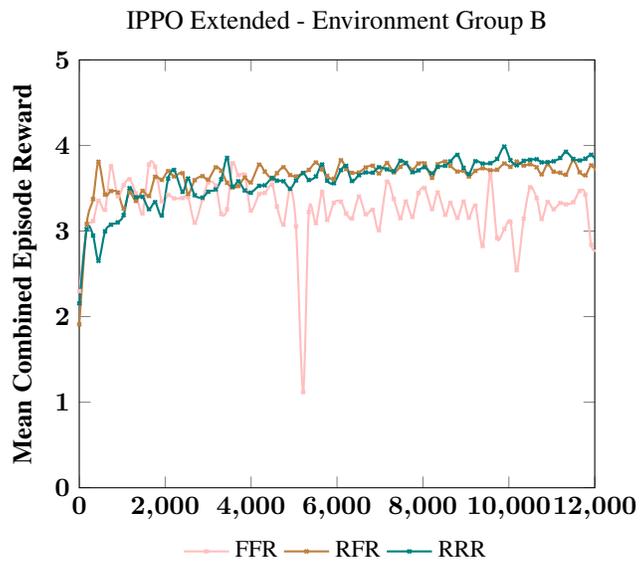

Our results in Figure \ref{fig:longexps}, show that IPPO agents continue to converge to the suboptimal strategy previously observed, despite the hyperparameter tuning and extended training duration. This indicates that the cause of this behaviour is likely neither insufficient training time nor unsuitable hyperparameter values. Rather, our results suggest that the convergence behaviour displayed is a direct result of the increased complexity of environments in Group B. 

\end{document}